\documentclass[journal=jacsat,manuscript=article]{achemso}

\usepackage{chemformula} 
\usepackage[T1]{fontenc} 

\def \bii{BiI$_{3}$} 
\def \Fref{Fig.~\ref}

\author{Valentina Gosetti}
\affiliation{Department of Mathematics and Physics, Universit\`a Cattolica, I-25133 Brescia, Italy}
\altaffiliation{Interdisciplinary Laboratories for Advanced Materials Physics (I-LAMP), Universit\`a Cattolica, I-25133
Brescia, Italy}
\alsoaffiliation{Department of Materials Engineering, KU Leuven, Kasteelpark Arenberg 44, 3001 Leuven, Belgium}

\author{Jorge Cervantes-Villanueva}
\affiliation{Institute of Materials Science (ICMUV), University of Valencia,  Catedr\'{a}tico Beltr\'{a}n 2,  E-46980,  Valencia,  Spain}

\author{Davide Sangalli}
\affiliation{Istituto di Struttura della Materia-CNR (ISM-CNR), Area della Ricerca di Roma 1, Monterotondo Scalo, Italy}

\author{Alejandro Molina-Sánchez}
\affiliation{Institute of Materials Science (ICMUV), University of Valencia,  Catedr\'{a}tico Beltr\'{a}n 2,  E-46980,  Valencia,  Spain}

\author{Vadim F. Agekyan}
\affiliation{St. Petersburg State University, St. Petersburg, 199034, Russia}

\author{Claudio Giannetti}
\affiliation{Department of Mathematics and Physics, Universit\`a Cattolica, I-25133 Brescia, Italy}
\altaffiliation{Interdisciplinary Laboratories for Advanced Materials Physics (I-LAMP), Universit\`a Cattolica, I-25133
Brescia, Italy}

\author{Luigi Sangaletti}
\affiliation{Department of Mathematics and Physics, Universit\`a Cattolica, I-25133 Brescia, Italy}
\altaffiliation{Interdisciplinary Laboratories for Advanced Materials Physics (I-LAMP), Universit\`a Cattolica, I-25133
Brescia, Italy}

\author{Selene Mor}
\affiliation{Department of Mathematics and Physics, Universit\`a Cattolica, I-25133 Brescia, Italy}
\altaffiliation{Interdisciplinary Laboratories for Advanced Materials Physics (I-LAMP), Universit\`a Cattolica, I-25133
Brescia, Italy}

\author{Stefania Pagliara}
\affiliation{Department of Mathematics and Physics, Universit\`a Cattolica, I-25133 Brescia, Italy}
\altaffiliation{Interdisciplinary Laboratories for Advanced Materials Physics (I-LAMP), Universit\`a Cattolica, I-25133
Brescia, Italy}
\email{stefania.pagliara@unicatt.it}

\title[An \textsf{achemso} demo]
 {Detection of a coherent excitonic state in the layered semiconductor \bii}%

\begin{document}



\begin{abstract}
The measurement and manipulation of the coherent dynamics of excitonic states constitute a forefront research challenge in semiconductor optics and in quantum coherence-based protocols for optoelectronic technologies. Layered semiconductors have emerged as an ideal platform for the study of exciton dynamics with accessible and technologically relevant energy and time scales. Here, we investigate the sub-picosecond exciton dynamics in a van-der-Waals semiconductor upon quasi-resonant excitation, and achieve to single out an incipient coherent excitonic state. Combining broadband transient reflectance spectroscopy and simulations based on many-body perturbation theory, we reveal a transient enhancement of the excitonic line intensity that originates from the photoinduced coherent polarization that is phase-locked with the interacting electromagnetic field. This finding allows us to define the spectral signature of a coherent excitonic state and to experimentally track the dynamical crossover from coherent to incoherent exciton, unlocking the prospective optical control of an exciton population on the intrinsic quantum-coherence timescale.

\end{abstract}

\section{Introduction}
The coherent interaction between photons and the matter induces a polarization that is transiently phase-locked with the electromagnetic field. In semiconductors, resonant laser pulses can be used to directly generate electron-hole pairs (excitons) \cite{Chen2016, Steinleitner2017,Selig2018}. In the coherent interaction regime, the exciton population refrains from scattering events. Moreover, the field-induced polarization may enable transient optical responses that are otherwise inaccessible. After the interaction with the laser pulse, many-body scattering processes lead to the dephasing of the field-induced polarization \cite{Cassette2014} and the decay of the exciton population \cite{Simbula2023}. Identifying the build-up of a coherent excitonic state \cite{Trovatello2020B} and tracing the crossover from coherent to incoherent exciton dynamics \cite{PhysRevB.62.16802} are the essential challenges towards the control of quantum coherence in functional materials of interest for the next-generation quantum technologies  \cite{Duong2017,Mueller2018, Wang2021,Montblanch2023}. 

The access to the early-time coherent regime in solid state materials is far from trivial from both an experimental and a theoretical viewpoint. 
The coherent coupling between light and excitonic transitions breaks rapidly  within up to a few tens of femtoseconds (fs) due to various dissipative channels such as exciton-exciton collisions, exciton-phonon scattering events~\cite{Selig2016,Chen2020,Cudazzo2020,Antonius2022,Lechifflart2023,Chan2023}, and the presence of defects or impurities \cite{Ruppert2017}.
To date, transient optical spectroscopy, which is the elective technique used by the scientific community studying the non-equilibrium physics in condensed matter, turns out to be not suitable for the study of this coherent regime.
This is better probed by multi-dimensional spectroscopy, a complex technique that allows to discriminate the dephasing mechanisms by isolating the homogeneous and the inhomogeneous broadening of the exciton linewidth, where the first represents the quantum coherent dynamics and the latter the scattering with defects and impurities \cite{Moody2015, Jakubczyk2019, Li2021}. Conversely, the measurement of coherent excitonic states by transient optical spectroscopy is still elusive as it demands the careful preparation of these states through a proper choice of the experimental parameters~\cite{Franceschini2023} and the establishment of unambiguous spectral signatures of the coherent excitonic state to be compared with theoretical predictions are mandatory.

From a theoretical perspective, the use of \textit{ab initio} approaches to study femtosecond exciton dynamics has been eluded due to the computational demand. Recent \textit{ab initio} developments model the generation and detection of (coherent) excitons~\cite{Merkl2019,Sangalli2021,Michael2021,dong_direct_2021, Cistaro2023}, the stability of excitons against non-equilibrium screening~\cite{Perfetto2020}, as well as exciton coherent dynamics and transport~\cite{cohen2023phonondriven}.
These developments aim to determine the fingerprints of non-equilibrium excitons in pump-probe experiments and provide a framework to distinguish signatures of (coherent) excitons from free-carrier populations. However, proper approaches to model the transition from the coherent to the non-coherent excitons, and to distinguish their signature in transient spectroscopy, are still missing. Few theoretical studies are focused on time-resolved ARPES, usually limited to models~\cite{Christiansen2019,Stefanucci2021}, with interesting ongoing developments to account for non-Markovian dynamics, and coupling with lattice degrees of freedom~\cite{perfetto_real-time_2023}.

In the current effort to identify coherent excitonic states and measure their decoherence time, low-dimensional materials arise as the ideal platform to explore ultrafast exciton dynamics. Excitons in layered materials have a large binding energy due to the reduced dielectric screening of the Coulomb interaction \cite{Chernikov2014}. For instance, 2D transition metal dichalcogenides (TMDCs) have been extensively investigated to understand exciton formation \cite{smejkal_time-dependent_2021,Trovatello2022,dong_direct_2021}, exciton coupling to coherent phonons \cite{Trovatello2020},  high-harmonic generation \cite{liu_high-harmonic_2017}, valley polarized exciton dynamics\cite{MolinaSanchez2017}, and formation of Floquet states by light dressing of excitons \cite{kobayashi_floquet_2023, Cunningham2019, Sie2015}. 

Another interesting layered family for the investigation of the exciton dynamics is represented by metal-halide semiconductors \cite{Wang2021}. Within this family, \bii\ hosts direct excitons in the visible range with even higher binding energy than that of the more commonly studied TMDCs both in two-dimension as well as in the bulk\cite{JorgeArxiv, Kaifu1988,Jellison1999}. Thanks to these peculiarities, \bii\ is triggering an increasing interest for the study of exciton scattering mechanisms with the environment, as recently shown by the study of the exciton coupling with coherent optical phonons\cite{Scholz2018,Mor2021,Mor2024}.
Literature reports numerous efforts to address the early-time regime of the excitonic resonance through transient optical spectroscopy, though without reaching consensus on the spectral signature of a coherent excitonic state. Different signs and energy shifts of the transient signal at the excitonic energy have been reported depending on the investigated system, and on the fluence and the photon energy of the pump pulse \cite{Ugeda2014, Chernikov2015, Pogna2016,Baldini2020} .

In this work, we resolve a coherent excitonic state beyond the laser pulse duration, by transient reflectance spectroscopy, through the successful choice of the experimental parameters for the preparation of this state, and we validate the spectral fingerprint of the coherent excitonic state by state-of-art \textit{ab-initio} real-time simulations. 
We choose the main exciton of \bii\ and we measure the photoinduced reflectivity variation upon quasi-resonant excitation of the lowest bright exciton transition and at low temperatures. The real-time simulations are based on the equation of motion for the density matrix, including excitonic effects, and simulate simultaneously the pump and probe effect on the system. The agreement between experiment and theory establishes unambiguously the signature of a coherent excitonic state on the transient optical spectrum.

\section{Experimental}

The study of the exciton dynamics of \bii\ has been carried out by a common transient reflectance (TR) experimental setup. 
Probe pulses are generated 
from a 270-fs-long pulse at 1030 nm passing through a 4-mm-thick YAG crystal and cover an energy window from 1.9 eV to 2.3 eV.
Pump pulses with tunable energy are generated by an optical parametric amplifier and then compressed, at 650 nm the bandwidth is $\Delta \lambda$ =12 nm. 
The instrumental response function, defined as the cross-correlation between the pump and the probe measured at the sample position, is limited by the temporal duration of the pump pulse and results 70 fs (see SI for further details).

\section{Results and discussion}
\begin{figure}
\includegraphics[width=1\columnwidth]{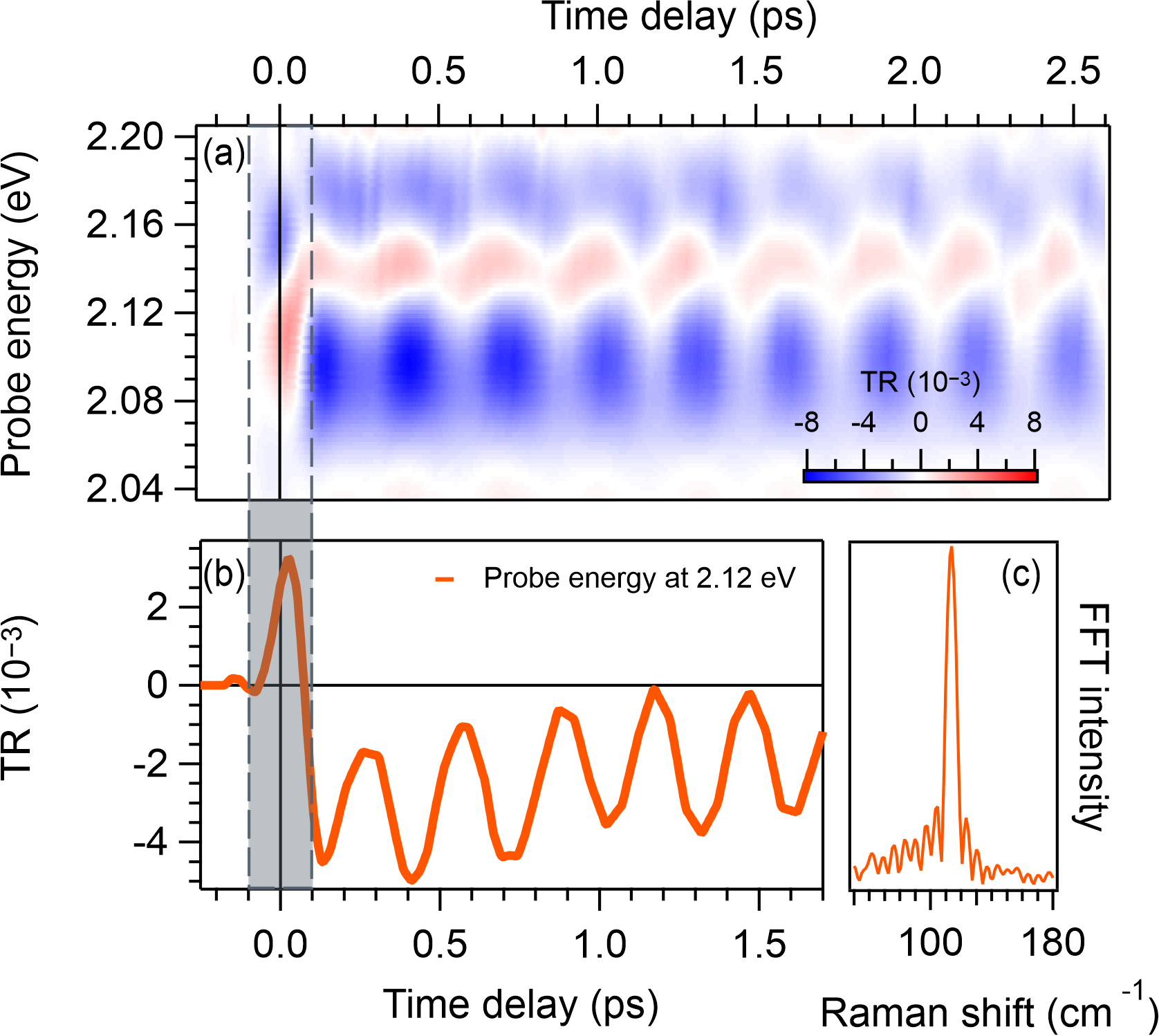}
\caption{ (a) TR of \bii\, upon quasi-resonant excitation (1.91~eV) at 50~K as a function of pump-probe time delay (bottom axis) and probe photon energy (left axis). (b) Time-resolved line cut at 2.12~eV of the TR map in (a). (c) FFT of the periodic modulation of the time-resolved linecut in (b).}
\label{Fig1}
\end{figure}

\Fref{Fig1}(a) shows the transient reflectance (TR) collected on a \bii\ single crystal as a function of pump-probe time delay (bottom axis) and probe photon energy (left axis). 
We choose a pump photon energy of 1.91~eV in order to minimize the excitation of quasi-free carriers to the conduction band continuum, in favor of the quasi-resonant generation of excitons that will be initially phase-locked with the absorbed photons. The sample temperature is set at 50~K. 
We identify two temporal regimes in the TR color map: one initiated at the pump pulse arrival and persisting up to approximately 100 fs, and a second one after 100 fs. In the early-time regime ($<$~100 fs), the TR signal is positive below 2.12~eV, and negative at higher energies. During the late-time regime ($>$~100 fs), the TR intensity is (i) negative in the spectral region below 2.12~eV, (ii) positive up to 2.16~eV, (iii) negative at energies higher than ca. 2.16~eV. Finally, at all time delays and at any probe photon energy, the TR intensity oscillates periodically, as highlighted by the horizontal line cut at 2.12~eV reported in \Fref{Fig1}(b). The Fast Fourier transformation (FFT) shown in \Fref{Fig1}{(c)} reports a Raman shift of 113.60 $\pm$ 0.02 cm$^{-1}$, which corresponds to the dominant $A_g$ phonon mode of \bii\, coherently generated by the pump pulse~\cite{Scholz2018,Mor2021}.

\begin{figure}
\includegraphics[width=1\columnwidth]{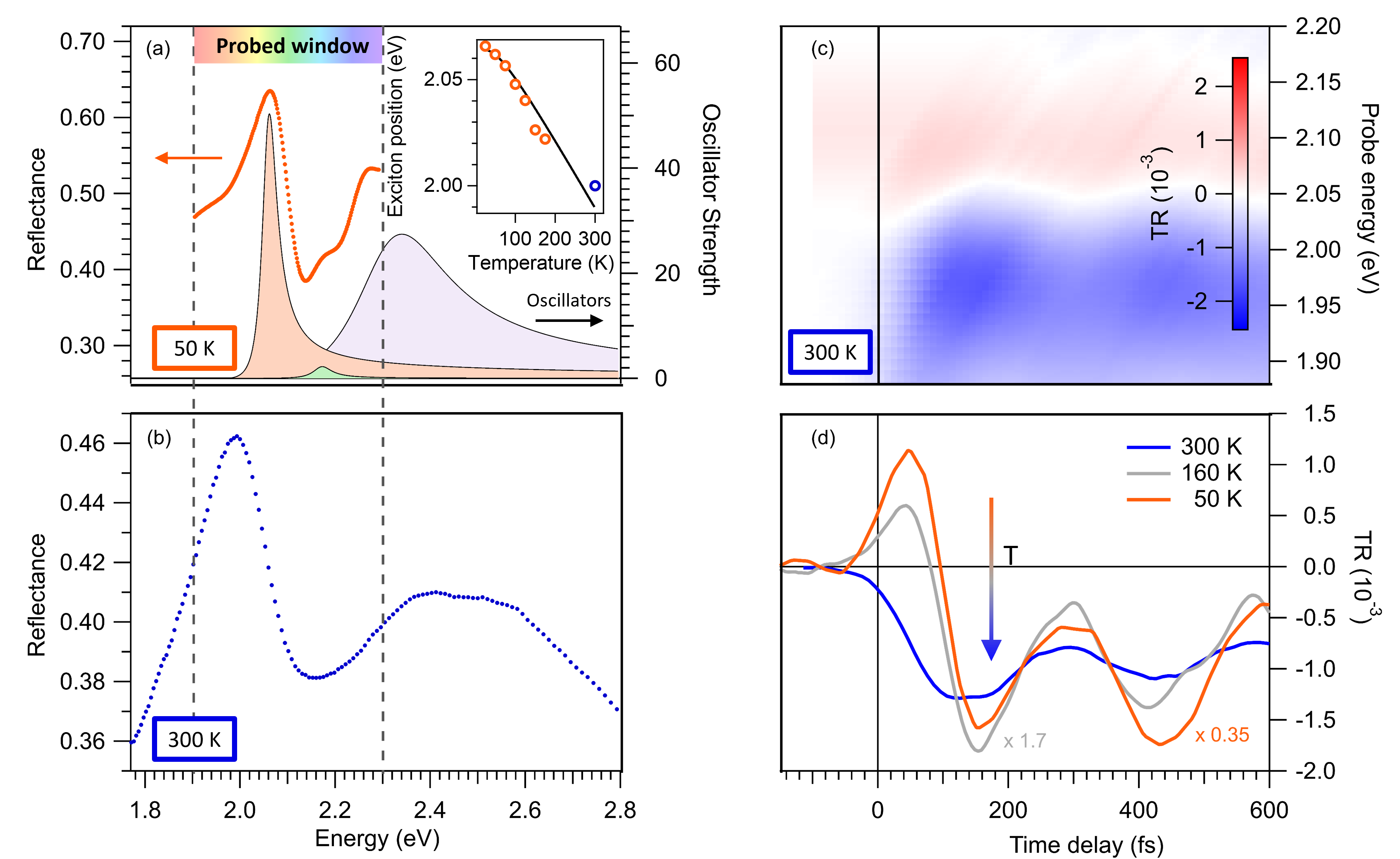}
\caption{(a) On the left axis, the equilibrium reflectance (orange markers) collected at 50 K by using the continuum probe. On the right axis, the three oscillators (orange, green, and violet shades) resulting from a Tauc-Lorentz analysis. The inset shows the temperature dependence of the first excitonic resonance measured with the continuum probe (orange markers). The data point at 300 K is extracted from the equilibrium reflectance collected by the UV/VIS spectrophotometer and shown in (b) (see S.I. for details) . (c)  TR map of \bii\ upon quasi-resonant pumping (1.91 eV) as a function of probe energy (right axis) and time delays (bottom axis) at 300 K. (d) Time-resolved line cuts at the energy where the intensity is minimum for three temperatures: 50 K, 160 K, and 300 K. The traces are normalized on the negative TR signal at 200 fs.}
\label{Fig2}
\end{figure}

To determine the energy region of the TR map connected to the dynamics of main excitonic resonance, we measure the equilibrium reflectance at 50~K in the TR energy region and we interpolate the spectrum by the sum of three Tauc-Lorentz oscillators: the first one at (2.06$\pm$ 0.01) eV accounts for the main lowest-energy bright exciton, and the other two at (2.17$\pm$ 0.01) eV and (2.31$\pm$ 0.01) eV for a set of excitons at higher energies (\Fref{Fig2}(a)). The latter are confirmed by the equilibrium reflectance collected at room temperature in a broader energy range by means of a spectrophotometer (\Fref{Fig2}(b)) (See S.I. for details).  For completeness, the inset of \Fref{Fig2}(a) shows the continuous shifting of the main excitonic resonance from (2.06$\pm$ 0.01) eV at 50 K to (2.00$\pm$ 0.01) eV at 300 K upon temperature increase.  
As reported in the literature, this main exciton has a Frenkel-like character \cite{KAIFU198861,Mor2021} and its calculated binding energy is 140 meV \cite{JorgeArxiv}. 
Thus, the TR intensity observed in \Fref{Fig1}(a) below 2.12 eV at 50~K is assigned to the photoinduced dynamics of the main exciton. 

To rationalize the drastic TR sign change from the early- to the late-time regime, we explore different environmental conditions of the exciton dynamics which mainly affect the dephasing process of any initial coherent excitonic state. We start with the study of the effect of the temperature on the TR map. \Fref{Fig2}(c) shows that the TR change, observed at low temperature, is completely absent in the TR map at 300 K, being the TR signal at the excitonic resonance negative for all time delays.
The TR response at the probe energy of the excitonic resonance, collected at different temperatures (T= 50 K, 160 K and 300 K)  and upon quasi-resonant pumping, confirms that the initial positive signal vanishes with the temperature while the negative bleaching of the late-time regime and the periodic oscillations due to the coherent phonons persist (\Fref{Fig2}(d)).
In this light, we can assert that the low temperature, reducing the scattering mechanisms with phonons \cite{Ashcroft}, is a necessary requirement for the observation of the early-time exciton dynamics.

\begin{figure}
\includegraphics[width=0.7\columnwidth]{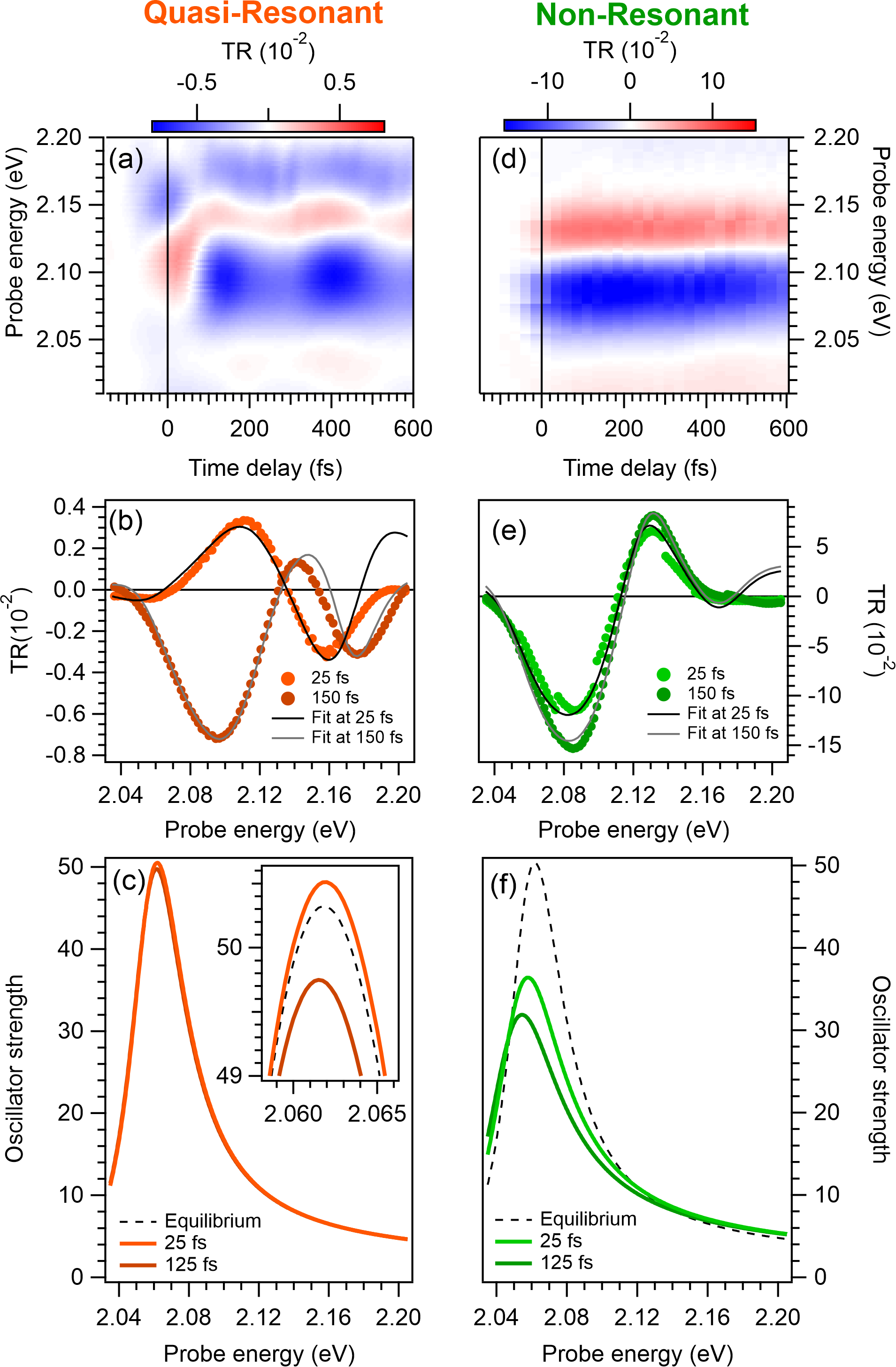}
\caption{TR of \bii\, at 50~K upon photoexcitation with 1.91~eV (a) and 2.79~eV (d) pump photons. Vertical line cuts, (b) and (e), of the color map in (a) and (d), respectively. Oscillator strength of the first exciton at equilibrium, 25 fs and 150 fs upon photoexcitaton with (c) 1.91~eV and (f) 2.79~eV pump photons. }
\label{Fig3}
\end{figure}

\Fref{Fig3}(a) singles out the early-time TR response at 50 K to introduce a more quantitative analysis that allows us to extract the photoinduced changes in both the oscillators parameters and the energy gap through differential fitting of the TR spectra at various pump-probe time delays. Two vertical line cuts of the TR map are shown as markers in \Fref{Fig3}(b) superimposed by the relevant best-fit curves (solid lines).
The positive TR in the early-time regime (25~fs) results from the combination of a blueshift and an intensity increase of the main oscillator.  
Conversely, the negative TR signal in the late-time regime (150~fs) originates from the bleaching of the oscillator strength accompanied by a redshift of the exciton energy (dark orange line in \Fref{Fig3}(c)).

We now examine the TR map at 50~K upon pumping with photons at 2.79~eV (\Fref{Fig3}(d)), which exceeds the energy of the optical gap of \bii. By that, we intend to 
enable the excitation of quasi-free carriers to the conduction band. As a consequence, unlike the quasi-resonant excitation condition, here carrier-carrier and exciton-carrier scattering processes are promoted. The relevant data are shown in \Fref{Fig3}(d). The time- and energy-resolved TR intensity preserves the same sign at any time delay: it is dominated by a negative TR in the region of the main excitonic resonance (below 2.12~eV) and a positive intensity at higher energies. Two vertical line cuts at distinct pump-probe time delays are shown \Fref{Fig3}(e) to highlight that upon off-resonant excitation the TR signal does not exhibits any drastic sign change over time. Rather, at any time delays the TR spectra results from both the bleaching and the redshift of the lowest-energy exciton oscillator (light and dark green traces in \Fref{Fig3}(f)). We note that these results are in line with the dynamics initiated upon quasi-resonant excitation and observed at late time delays.
 
The comparison between the TR under quasi- and off-resonance excitation highlights that the exciton dynamics occurring at early time delays is a unique feature. In fact, its solely appears when the laser pump pulse coherently excites excitons, while the excitation of quasi-free carriers in the continuum is kept negligible. Moreover, we recall that the low temperature minimizes, as previously discussed, the scattering processes with thermal phonons.
All these elements, together with the short timescale of the relevant dynamics, agree that the early-time regime is the manifestation of a coherent excitonic state.

\begin{figure}
\includegraphics[width=0.96\columnwidth]{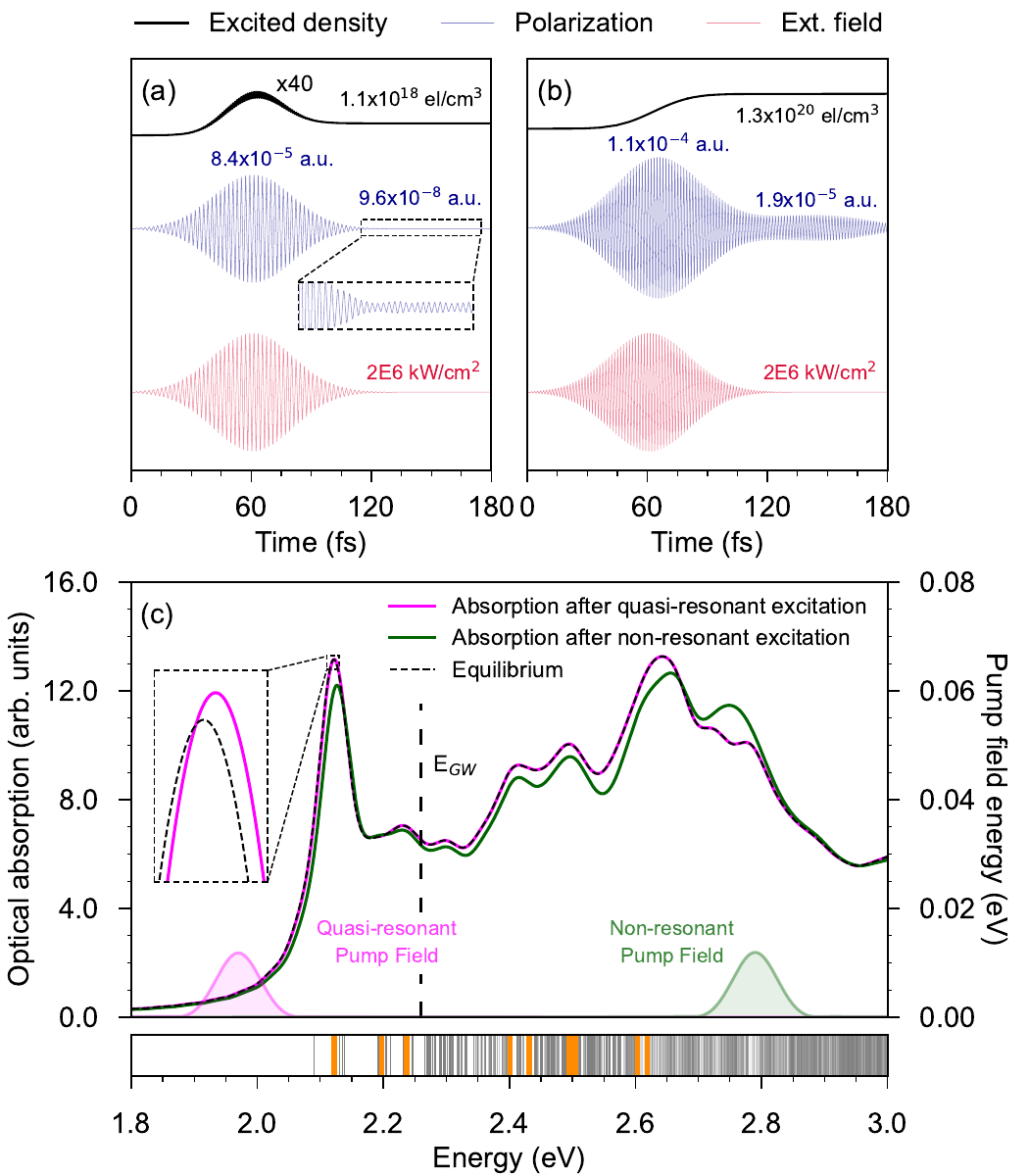}
\caption{Equilibrium and non-equilibrium optical properties of BiI$_{3}$ from first-principles calculations. (a-b) Excited density, polarization and pump field for (a) the pump at 1.97 eV and (b) the pump at 2.79 eV. In (a) the excited density has been multiplied by a factor of 40 to be easily observed, whereas the inset shows the non-zero polarization after the pump field. The values for the residual total excited density, the maximum polarization, the residual polarization and the pump field intensity are displayed. (c) Optical absorption spectrum for the equilibrium and non-equilibrium cases. The dashed vertical line represents the direct gap and the shaded areas the different pump fields. The inset in (c) shows the increase in intensity of the excitonic peak for the quasi-resonant case, while the inset at the bottom of the figure shows the bright (yellow) and dark (gray) excitons.
}
\label{Fig4_theo}
\end{figure}

In order to confirm this hypothesis, we compute the optical properties of BiI$_{3}$ both at equilibrium and out-of-equilibrium with an ab-initio approach based on many-body perturbation theory. First, the equilibrium absorption is computed via the GW+BSE scheme. This leads to a direct gap of 2.24 eV and a series of excitonic peaks, with the low energy in-plane spectrum dominated by a bright exciton at 2.12 eV. Although temperature effects are not considered in these calculations, the theoretical excitonic peak is in good agreement with the low-temperature measurement shown in Fig. \ref{Fig2}(a). Then, we perform real-time simulations based on the equation of motion for the density matrix, $\rho(t)$ \cite{Sangalli2021}. In the simulation, we need to include two key ingredients: (i) the long-range screened electron-hole Coulomb interaction to account for the generation of coherent excitons, and (ii) both the pump and the probe laser pulses, to model the optical signal accounting for the coherent polarization, i.e. the coherent excitonic state, induced at early times. Further details on the numerical simulations are given in the SI.

We want to investigate how the non-equilibrium optical spectrum changes if excitons or carriers are injected into the material. Thus, we consider the two pumping condition used experimentally: (1) quasi-resonant case with a pump pulse with a detuning of 150 meV with respect to the theoretical excitonic peak for the quasi-resonant case ($\omega_P=1.97$ eV), and (2) off-resonance case with a pump pulse in the continuum of the absorption spectrum ($\omega_P=2.79$ eV). The corresponding results are depicted in Fig.~\ref{Fig4_theo}, where in panel Fig.~\ref{Fig4_theo} (a-b) the output of the simulations including only the pump pulse are shown, with the field of the pump laser pulse, the induced polarization, and the total excited population density.

In the quasi-resonant case displayed in Fig.~\ref{Fig4_theo}(a), both population and polarization reach a maximum slightly delayed compared to the maximum of the pump pulse, and then the decay to a much smaller value. This decay is due to the detuning between the pump frequency and the excitonic peak, and can be described in terms of a large portion of virtual coherent states which do not survive beyond the pump laser pulse. The residual population and polarization instead corresponds to real states excited in the material, and exists because the long-range Coulomb interaction allows absorption below the direct gap. This residual polarization oscillates at a frequency which exactly matches the excitonic frequency, thus indicating that the residual population density corresponds to a population of coherent excitons~\cite{Perfetto2019b,Sangalli2021}. 

The situation is significantly different when the system is excited by a laser pulse with frequency well above the fundamental gap, Fig.~\ref{Fig4_theo}(b). In this case, the excited density reaches a maximum at the end of the pump pulse and does not decay subsequently. This means that the laser pulse is completely absorbed and therefore only real states are excited. Since the pump frequency lays in the continuum, these states can be described in terms of free carriers, while the associated  polarization decays quickly because of the so called free polarization decay. In the simulations, this does not translate into an actual dephasing, but rather to destructive interference between the many frequencies excited (the residual polarization is due to the fact that numerically we have a finite number of frequencies only). Therefore, we have a population of coherent carriers, although here the coherent nature of the states does not show up in the polarization. Experimentally we expect such coherences to decay due to dissipative mechanisms. In any case, the free carriers coherences do not have any impact on any result of the simulations (see also SI).

Now, we consider the non-equilibrium optical signal in these two configurations and compare it with the equilibrium absorption spectrum, as shown in Fig.~\ref{Fig4_theo}(c). To this end, we use a probe laser pulse which arrives 25 fs after the maximum of the pump pulse, as in the experimental setup. To obtain the optical signal, we define the probe only polarization $P_p(t)=P_{Pp}(t)-P_P(t)$, e.g. the result of 
subtracting the polarization induced in presence of the pump-only, from the polarization induced in presence of both  pump and probe. The two non-equilibrium spectra obtained have a striking difference: the signal due to the excitonic population leads to an enhancement of the excitonic peak, while the signal due to the carriers results in a bleaching of the same peak. This explains the opposite signal measured experimentally at 25 fs in Fig.~\ref{Fig3}. Moreover, we also notice that the signal induced by the injected carriers affects the absorption spectrum over a large energy range, whereas the excitonic population only affects the signal nearby the excitonic peak. The signal in the non-resonant case is not affected by the coherent nature of the states. Indeed, the same signal is obtained both changing the pump-probe delay or considering a population of non-coherent carriers. Instead, the enhancement of the excitonic peak is largely due to the coherent nature of the excitonic state. By varying the time delay between the pump and the probe pulse, the excitonic peak intensity displays time oscillations with a frequency matching twice the excitonic frequency. This is known as 
the Dynamical Franz-Keldish effect \cite{Pedersen2016} and is caused by the induced polarization. The signal shown in Fig.~\ref{Fig4_theo}(c) is obtained by averaging over two probes with opposite relative phase which washes out the time dependent oscillations (see SI for details).

\begin{figure}
\includegraphics[width=0.9\columnwidth]{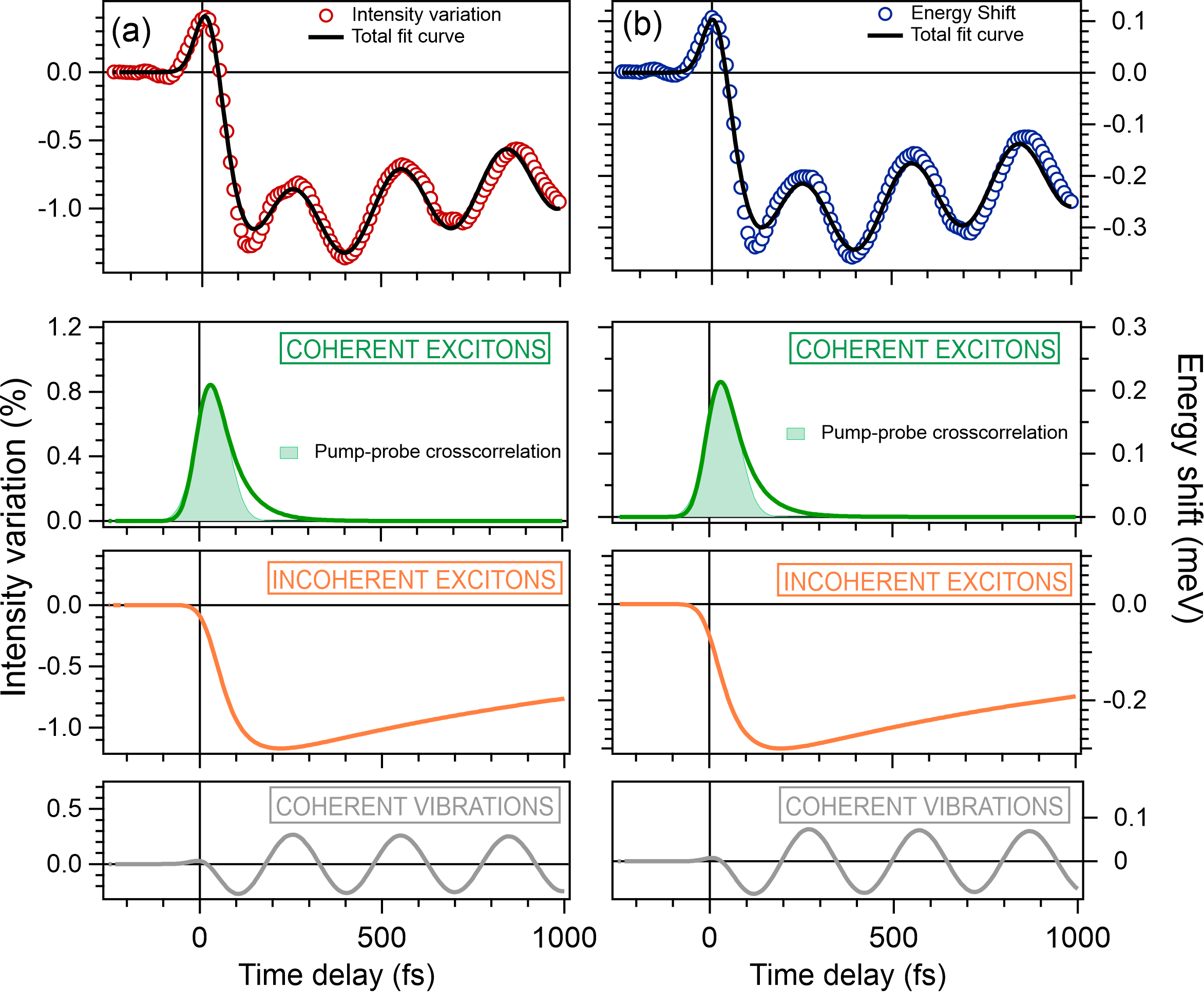}
\caption{(a) Time-resolved intensity variation (red markers) of the excitonic oscillator and relevant fit (solid black line). (b) Time-resolved energy shift (blue markers) of the excitonic oscillator and relevant fit (solid black line). The coherent excitons, incoherent excitons, and coherent vibrations contributions are shown as green, orange, and gray curves, respectively, for both (a) and (b).}
\label{Fig5}
\end{figure}
In order to trace the dynamical crossover from coherent to incoherent excitonic state, we follow the intensity (\Fref{Fig5}(a)) and energy (\Fref{Fig5}(b)) variation of the excitonic oscillator (\Fref{Fig5}(a)) as a function of time, as obtained from differential fitting of the TR spectra. The intensity variation is positive within approximately 200 fs, then evolves into negative before vanishing on a several ps timescale (not shown). The excitonic resonance shows an initial energy blueshift on the same timescale as the TR intensity increase, followed by a redshift that recovers in a few-ps. 
Both the intensity and the energy are periodically modulated at the frequency of the A$_g$ phonon mode, due to the exciton-phonon coupling \cite{Mor2021}.

To decouple the contributions to the coherent-to-incoherent crossover dynamics, we \textit{globally} fit the time-resolved intensity variation and the energy shift of the excitonic oscillator through the sum of three terms (shown separately in \Fref{Fig5}), convoluted with the pump-probe cross-correlation (accounting for the resolution time, see SI).
The early-time component (green curve) is associated to the dephasing dynamics of the coherent excitonic state and shows a decay time of $(60 \pm 10)$~fs, resulting in a dynamics that outlives the regime of pump-probe temporal overlap 
(green shade, the full width at half maximum is 70~fs).   
The orange lines show a delayed minimum, followed
by a slowly-recover with a component dynamics, one with a time constant of (1.2 ± 0.1) ps and the second
longer than the probed window. 
Notably, the minimum is delayed by the same time as the dephasing of the excitonic coherent state, tracing the crossover from coherent to incoherent exciton population. This is followed by the relaxation and recombination of the incoherent exciton population. Eventually, the cosine-like modulation (grey line) of both the exciton intensity and energy with a period of $(300 \pm 10)$~fs (i.e. ($111 \pm 5)~cm^{-1}$) is indicative of exciton-phonon coupling persisting until coherent optical phonons dephase on a long timescale of several ps by virtue of the low temperature of the system.

We note that the estimated dephasing time of $(60 \pm 10)$~fs being much lower than the depopulation one (few nanoseconds), determines a homogeneous exciton broadening of about $(11 \pm 5)$~meV. The intrinsic excitonic linewidth, estimated by the equilibrium reflectance measurements, is $(15 \pm 5)$ meV (see SI for details) which is in good agreement with the homogeneous broadening estimated by the dephasing time. This result confirms the absence of defects or impurities, responsible of the inhomogeneous broadening, in our \bii\   
sample and indicates that a further requirement to observe the coherent excitonic state
is the presence of an exciton with a linewidth close to the homogeneous limit.

\section{Conclusion}

\subsection*{Acknowledgements}
D.S. acknowledges funding from MaX ``MAterials design at the eXascale'' (Grant Agreement No. 101093374) co-funded by the European High Performance Computing joint Undertaking (JU) and participating countries, and PRIN Grant No. 20173B72NB funded by MIUR (Italy). A.M.S. and J.C.V. acknowledge the funding of Ministerio de Ciencia e Innovación, which is part of Agencia Estatal de Investigación (AEI), through the project PID2020-112507GB-I00 QUANTA-2DMAT (Novel quantum states in heterostructures of 2D materials), the Generalitat Valenciana through the Grant PROMETEO/2021/082 (ENIGMA) and the projects SEJIGENT/2021/034 and MFA/2022/009 . J. C.-V. acknowledges the Contrato Predoctoral Ref. PRE2021-097581. A. M.-S. acknowledges the Ram\'on y Cajal programme (grant RYC2018-024024-I; MINECO, Spain). This study forms part of the Advanced Materials programme and was supported by MCIN with funding from European Union NextGenerationEU (PRTR-C17.I1) and by Generalitat Valenciana.

\begin{suppinfo}

The Supporting Information is available free of charge at xxx.

Crystal growth; Optical set-up; Data processing for the correction of the  supercontiuum white-light temporal
chirp; \bii\, equilibrium reflectance and more details on the theoretical calculations.
\end{suppinfo}

\bibliography{BiI_bib.bib}

\end{document}